\documentclass{article}

    \usepackage[preprint]{neurips_2020}

\usepackage[utf8]{inputenc} % allow utf-8 input
\usepackage[T1]{fontenc}    % use 8-bit T1 fonts
\usepackage{hyperref}       % hyperlinks
\usepackage[ruled,vlined]{algorithm2e}
\SetAlFnt{\small}
\SetAlCapFnt{\large}
\SetAlCapNameFnt{\large}
\usepackage{graphicx}

\usepackage{titlesec}
\titlespacing*{\section}{0pt}{1.1\baselineskip}{\baselineskip}
\titlespacing*{\subsection}{0pt}{1.1\baselineskip}{\baselineskip}
\titlespacing*{\subsubsection}{0pt}{1.1\baselineskip}{\baselineskip}

\usepackage{placeins}

\usepackage{url}            % simple URL typesetting
\usepackage{booktabs}       % professional-quality tables
\usepackage{amsfonts}       % blackboard math symbols
\usepackage{nicefrac}       % compact symbols for 1/2, etc.
\usepackage{microtype}      % microtypography

\title{BumbleBee: A Transformer for Music}

\author{%
  Maria Juliana.~Quintero \\
  University of Toronto\\
  
  \And
   Lucas Fenaux \\
   University of Toronto\\
  
}

\begin{document}

\maketitle

\begin{abstract}
  We will introduce BumbleBee, a transformer model that will generate MIDI music data . We will tackle the issue of transformers applied to long sequences by implementing a longformer generative model that uses dilating sliding windows to compute the attention layers. We will compare our results to that of the music transformer and Long-Short term memory (LSTM) to benchmark our results. This analysis will be performed using piano MIDI files, in particular , the JSB Chorales dataset that has already been used for other research works (Huang et al., 2018)
\end{abstract}

\section{Introduction}
Generative models have progressed exponentially in the last few years, with multiple applications within text, speech and image generation. One of the main objectives in generative modelling is to capture different aspects of the data and to generate new instances indistinguishable from the true data.

Similarly, the usage of transformers, specifically,  self-attention, has been applied to different tasks such as reading comprehension, abstractive  summarization, textural entailment and learning task-independent sentence representation.  Self-attention is an attention mechanism which relates different positions of a single sequence in order to compute a representation of the sequence.

Music is inherently relative and self-referring with the presence of cycles and repetitions (for example the ABA format where a certain part is replayed a second time). It would then make sense to use models such as RelativeTransformers to generate music as their mathematical process covers all those aspects.
\section{Related Works}

Some data-driven approach to this problem include  DeepBach (Hadjeres et al., 2017) and Coconet (Huang et al., 2017) which use Gibbs sampling to produce notes in the style of Bach chorales, MidiNet (Yang et al., 2017) and MuseGAN (Dong et al., 2018) which use generative adversarial networks, MusicVAE (Roberts et al., 2018) and HRNN (Wu et al., 2019) which use hierarchical recurrent networks. Sequence models have been the standard choice for modeling music, such as using Long Short Term Memory networks (e.g., Eck \& Schmidhuber, 2002; Liang, 2016; Oore et al., 2018), to bidirectional LSTMs (e.g., Hadjeres et al., 2017.

Huang et al., (2018) presents the relative transformer and has become the current state of the art in music generation with transformers . It uses relative self-attention in order to capture and re-create relative musical patterns. It is very valuable as it does not suffer as much as LSTMs from long-sequence loss of attention and therefore performs much better when trained on long music sequences.

Dhariwa et al., (2020) present , a model that generates music with singing in the raw audio domain. It tackles the long context of raw audio using a multiscale VQ-VAE to compress it to discrete codes, and modeling those using autoregressive Transformers. The model has been used to generate high-fidelity and diverse songs with coherence up to multiple minutes.

\section{Method}
\subsection{Background} \label{3.1}
Our goal with this project is to be able to reproduce the results observed with the RelativeTransformer presented in Huang et al., (2018) as well as compare them with our own Long-Short Term Memory implementation.   The LSTM would serve as a baseline to compare our findings to as LSTMs have been a staple of signal processing as well as music generation. 

We then implemented the longformer, which is  meant  to improve the memory and time complexity issues that follow long sequences inputs into the transformer, reducing the implementation from O(ns×ns) to O(ns×w), with ns being the sequence length and w being the average window size.  The technique we will be using is dilating sliding windows and implementing some of the code found in this library LongFormers. 

We are using as our base code the music transformer implementation we found on github, with our own modifications allowing the code to run in our environment and reducing the data size. See Appendix: Modifications To The Original Code for details on the changes made.
\subsection{Music Transformer Architecture} \label{3.2}
As shown in Appendix: Figure 1, our model first takes the inputs and creates embeddings that are used for creating the positional encoding. Once this is completed, combining with the mask, the x inputs go through the DecoderLayer, which is define by the light blue squares, which is where the relative self-attention (see Appendix: Figure 2) and the Feed Forward layers lie. In our model, we have 4 DecoderLayers, thus this loop repeats four times and finally get normalized and goes through a final linear layer before produce the final output.
\subsection{Longformer Implementation} \label{3.3}
We changed the multihead attention from using relative attention to using a sliding window implementation (see Appendix Figure 3), in which we abstracted code from  \href{https://github.com/huggingface/transformers/tree/master/src/transformers/models/longformer}{\textit{HuggingFace’s Longformer}} and implemented the necessary adjustment for it to work within our module.  Below we have our unique changes we made to the forward method , any code we did not precisely show can be found in the LongformerSelfAttention class. 

\newpage
\begin{algorithm}[H]
\SetAlgoLined
\KwResult{Returns attention outputs}
 initialization\;
 same as MultiHeadAttention\;
 $self.one\_sided\_attn\_window\_size = attention\_window // 2$ \;
 Forward Pass \;
  \eIf{not self.relative\_pos}{
   $queries = queries / \sqrt{s}$ \# where $s =$ head dimension\;
   \# see LongformerSelfAttention for full implementation \;
   Chunk Queries and Keys into overlapping blocks of size w and overlap of size $\frac{1}{2}$ \;
   Multiply the blocks \;
   Mask out the diagonals \;
   Pad local attention probabilities by applying softmax to attention scores\;
   Apply dropout \;
   Get attention output by chunking V and attn probabilities into overlapping blocks of size w and overlap of size $\frac{1}{2}$ \; 
    Multiply the blocks \;
    use linear layer to recombine heads of the attention output \;
    \Return { attention output }}{
   complete regular MultiheadAttention as specified\;
  }
 \caption{LongMultiheadAttention Implementation}
\end{algorithm}

\subsection{Optimization: Early stopping}
We notice that our model would be running for at least 6-8 hours, however that the validation loss would increase towards the end of training, which led us to implement the early stopping condition. Since the validation error fluctuates because of the Adam optimizer use, we set our early stopping count to 200 batches in the validation batch. 

\subsection{Optimization: Initialization of trainable parameter weights}
As suggested in the paper by Huang et al., (2020) we applied Xavier initialization for all parameters excluding input embeddings. Use Gaussian initialization $N (0, d ^{ - \frac{1}{2}} )$ for input embeddings where d is the embedding dimension. We implemented this optimization to prevent the effects of large updates from
instability in the Adam optimizer coupled with increased
learning rate which tend to lead to gradient vanishing through layer normalization as describe in the paper.

\section{Experiments} 
For our experiments, we decided to observe how different models/optimization combinations performed in order to both validate the results observed in the Music Transformer paper and see if we could find a way to improve those results. In order to do so, we ran the music transformer as in \ref{3.2} , the longformer (\ref{3.3}) and finally a regular LSTM network to serve as a baseline.
To perform our experiments, we used the Piano-e-Competition dataset.
\\
We decided to train every combination of model/optimization in order to get as much data as possible as well as to observe how our optimizations affected the performance of each model.

\begin{table}
  \caption{Validation NLL for the Music Transformer using Piano-e-Competition dataset, with event-based representation with lengths L = 1024.}
  \label{music transformer}
  \centering
  \begin{tabular}{lll}
    \toprule
    % \multicolumn{2}{c}{Part}                   \\
    \cmidrule(r){1-2}
    Model Variation     & Validation NLL \\
    \midrule
    Music Transformer with no optimizations & 3.04    \\
    Music Transformer with xavier initialization     & 3.03    \\
    Music Transformer with early stopping      & 3.04 \\
    Music Transformer with both xavier initialization and early stopping    & 2.98 \\
    \bottomrule
  \end{tabular}
\end{table}

\begin{table}
  \caption{Validation NLL for the Longformer using Piano-e-Competition dataset, with event-based representation with lengths L = 1024.}
  \label{longformer_val}
  \centering
  \begin{tabular}{lll}
    \toprule
    % \multicolumn{2}{c}{Part}                   \\
    \cmidrule(r){1-2}
    Model Variation     & Validation NLL \\
    \midrule
    Music Longformer with no optimizations & 3.83    \\
    Music Longformer with xavier initialization     & 3.86    \\
    Music Longformer with early stopping      & 3.83 \\
    Music Longformer with both xavier initialization and early stopping    & 3.83 \\
    \bottomrule
  \end{tabular}
\end{table}

We can see in the Music Transformer's table below that while each optimization by itself provides minimal to non-existent improvement on the model's performance, we do get a more significant improvement when using both of the optimization methods together. As mentioned in our explaination of our optimization techniques, we can see that by implementing both techniques, we were able to stabilize the updates performed by the adam optimizer, allowing us to achieve a lower validation loss.

However, when we look at the table for the Longformer, we see a different picture before us. We find that there is no difference between not using any optimizations, using early stopping or using both early stopping and xavier initialization. We also find that the model actually performs worse when we only use xavier initialization by itself. This actually could be due to the xavier initialization helping the model reach is minima faster and without early stopping, the model ends up 'sabotaging' some of its earlier training and ends with a slightly worse performance. 

 The Music Transformer performs significantly better than the Longformer accross the board. This follows what we said in the introduction, that since music is inherently relative, the use of relative attention provides a stronger advantage for learning. The Longformer only used regular self-attention and therefore didn't benefit from this advantage, however, the use of the dilated sliding window did noticeably reduce training time (by roughly 30\% to 50\%). So it then becomes a matter of training speed versus training quality. 

When we look at our LSTM as a baseline, we can see that both the Music Transformer and Longformer before significantly better than the LSTM. This is both due to the Transformer and Longformer being more powerful models as well as the sequence length being a bit long for an LSTM (See Appendix: Extra Tables).

When listening to sound generated by the Music Transformer, while it's obviously not Mozart, we hear the model try different things, like holding certains keys for a while, or playing multiple keys that sound good together. When comparing with both optimizations and without optimizations, we can actually hear a clear difference. The one without optimizations tends to prefer more aggressive tones and the use of multiple keys at once. On the other hand, the one with both optimizations tends to prefer a less aggressive tone and seems to sometimes generate a piano scale which sounds quite good. 

For future experiments, we would combine the relative attention with our sliding window implementation, so that training time would reduce while maintaining the advantages relative attention provides for music, which might allow us to increase our model size which will likely increase performance.  Furthermore, we had omitted global attention and focused our Longformer to use only local attention, and thus for future experiments we would implement global attention and see if that would improve performance.
\\

\section{Conclusion }
In conclusion, we demonstrates that the relative transformer model as described by  Huang et al., (2018) and apply our own modifications to see if we can optimize the current model and found that through both xavier initialization and early stopping our model outperforms all other variations. We found that our Longformer was outperformed by the Relative Transformer due to the fact that music is inherently relative and that our Longformer was restrained to just local attention.
\footnote{For detailed contributions see 7.3 Appendix: List of Contributions}

\newpage

\section{References}
\small

[1] Cheng-Zhi Anna Huang, Ashish Vaswani, Jakob Uszkoreit, Ian Simon, Curtis Hawthorne, Noam Shazeer, Andrew M. Dai, Matthew D. Hoffman, Monica Dinculescu, and Douglas Eck. Music transformer: Generating music with long-term structure. In Seventh International Conference on Learning Representations, 2018 

[2] Hadjeres, G., Pachet, F., and Nielsen, F. Deepbach: a steerable model for bach chorales generation. In International Conference on Machine Learning, pp. 1362–1371. JMLR. org, 2017.

[3] Huang, C. A., Cooijmans, T., Roberts, A., Courville, A. C., and Eck, D. Counterpoint by convolution. In International Society for Music Information Retrieval Conference, pp. 211–218, 2017.

[4] Yang, L., Chou, S., and Yang, Y. Midinet: A convolutional generative adversarial network for symbolic-domain music generation. In International Society for Music Information Retrieval Conference, pp. 324–331, 2017.

[5] Dong, H.-W., Hsiao, W.-Y., Yang, L.-C., and Yang, Y.-H. MuseGAN: Multi-track sequential generative adversarial networks for symbolic music generation and accompaniment. In Thirty-Second AAAI Conference on Artificial Intelligence, 2018.

[6] Roberts, A., Engel, J., Raffel, C., Hawthorne, C., and Eck, D. A hierarchical latent vector model for learning longterm structure in music. In International Conference on Machine Learning, pp. 4364–4373, 2018. 

[7] Wu, J., Hu, C., Wang, Y., Hu, X., and Zhu, J. A hierarchical recurrent neural network for symbolic melody generation. IEEE Transactions on Cybernetics, 2019.

[8] Prafulla Dhariwal, Heewoo Jun, Christine Payne, 
Jong Wook Kim, Alec 
Radford, and Ilya Sutskever. Jukebox: A generative model for music.
https://cdn.openai.com/papers/jukebox.pdf, 2020. 

[9] Xiao Shi Huang, Felipe Perez, Jimmy Ba, and Maksims Volkovs. 2020. Improving transformer optimization through better initialization. In Proceedings of Machine Learning and Systems 2020, pages 9868–9876.

[10] Iz Beltagy, Matthew E. Peters, Arman Cohan. 2020. Longformer: The long-document transformer. arXiv
preprint arXiv:2004.05150

\newpage
\section{Appendix}
\subsection{ Extra Tables}
\FloatBarrier
\textbf{LSTM Table}
\begin{table}[h]
  \caption{Validation NLL for the LSTM using Piano-e-Competition dataset, with event-based representation with lengths L = 1024.}
  \label{longformer_vak_bis}
  \centering
  \begin{tabular}{lll}
    \toprule
    % \multicolumn{2}{c}{Part}                   \\
    \cmidrule(r){1-2}
    Model Variation     & Validation NLL \\
    \midrule
    Music LSTM with no optimizations & 5.25    \\
    Music LSTM with xavier initialization     & 5.25    \\
    Music LSTM with early stopping      & 5.25 \\
    Music LSTM with both xavier initialization and early stopping    & 5.24 \\
    \bottomrule
  \end{tabular}
\end{table}
\FloatBarrier

\subsection{Figures}
\begin{center}
\includegraphics[scale=0.4]{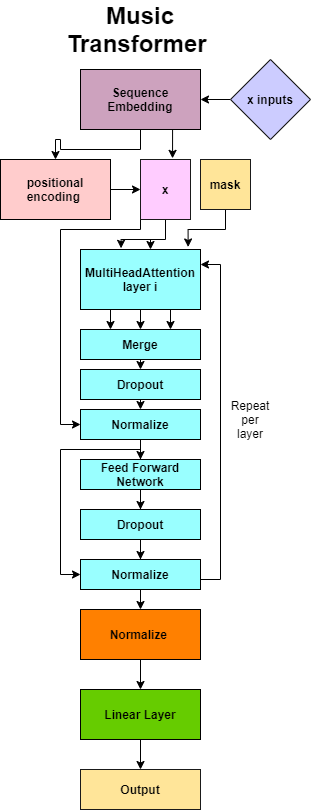}

\textbf{Figure 1} Music Transformer Architecture

\includegraphics[scale=0.25]{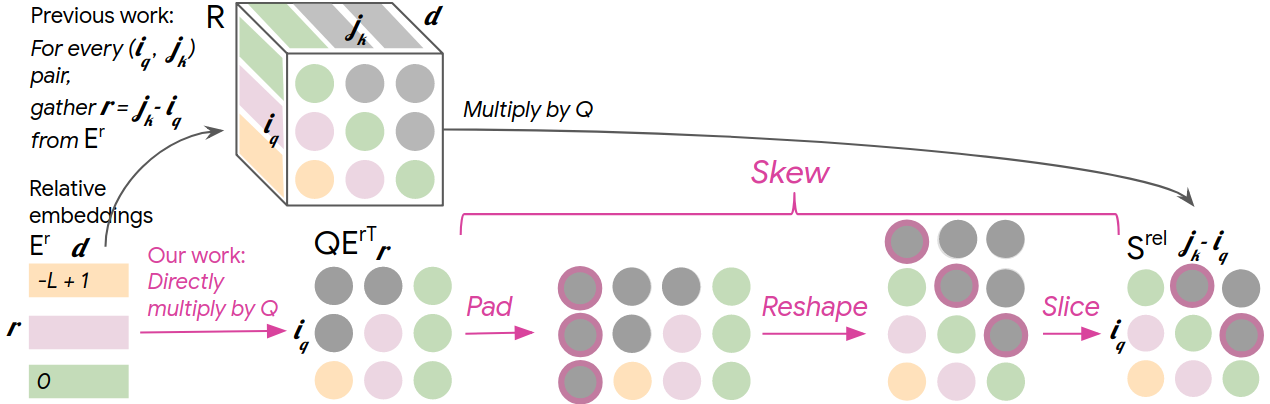}

\textbf{Figure 2} Huang et al. (2018) Relative Transformer Architecture

\includegraphics[scale=0.2]{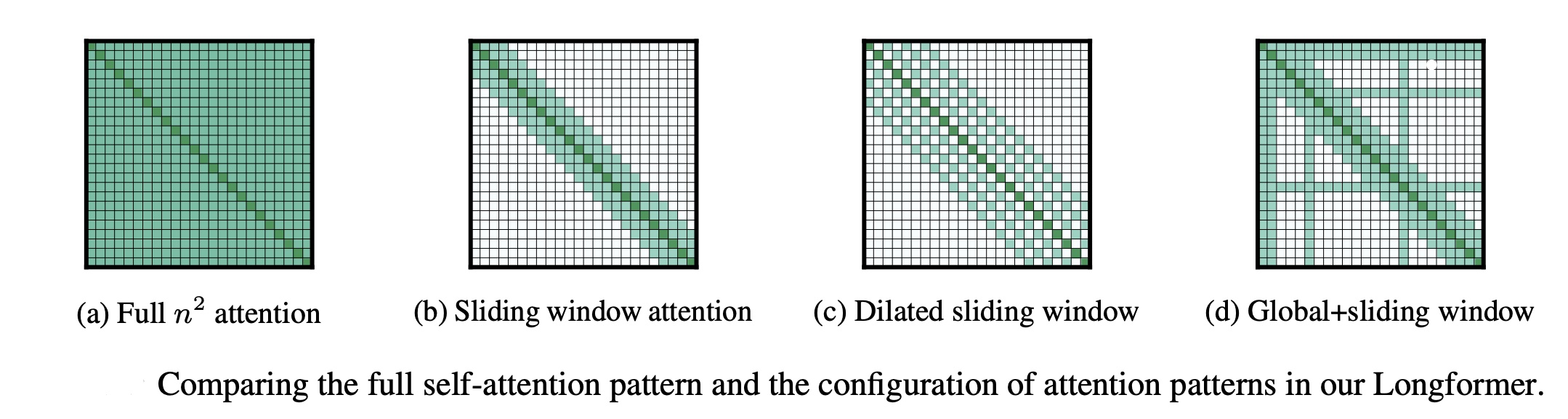}

\textbf{Figure 3} Beltagy et al. (2020) LongFormer Self-Attention Diagram

\end{center}

\subsection{Modifications To The Original Code}
In order to run our experiments, we re-used some of the code that was used in the original music transformer. This includes their implementation of the music transformer as well as their preprocessing pipeline and their training loop.
While the preprocessing pipeline and the transformer worked fine bar some small import and directory issues, there was several typos/bugs in the training loop that had to be fixed. On top of that, we had to revamp the model saving/loading process in order to use their music generation script. To view the final code used, check our repository on GitHub, \href{https://github.com/quinte22/bumblebee}{\underline{BumbleBee}}

\subsection{List of Contributions}
\subsubsection{Juliana's Contributions}
\begin{itemize}
    \item Implemented Longformer code
      \item Implemented Optimization code
      \item Wrote Method section, edited Experiment section
  \item Wrote abstract, conclusion, introduction and Appendix
      \item Co-Wrote related work

\end{itemize}

\subsubsection{Lucas' Contributions}
\begin{itemize}
    \item Fixed  music transformer code
    \item Implemented LSTM code
    \item Ran all experiments, music generation and logged their results
    \item Wrote Experiment section, edited Method section
    \item Co-Wrote related work
\end{itemize}
\end{document}